# Molecular Understanding of the Effect of Hydrogen on Graphene Growth by Plasma-Enhanced Chemical Vapor Deposition

Shiwen Wu[1,+], Dezhao Huang[2,+], Haoliang Yu[1], Siyu Tian[1], Arif Malik[1], Tengfei Luo[2,*], Guoping Xiong[1,*]

[1.] Department of Mechanical Engineering, The University of Texas at Dallas, Richardson, Texas 75080, USA

[2.] Department of Aerospace and Mechanical Engineering, University of Notre Dame, Notre Dame, Indiana 46556, USA

[+] These authors contributed equally to this work.

* Corresponding authors: Tengfei Luo: tluo@nd.edu; Guoping Xiong: guoping.xiong@utdallas.edu

# Abstract

Plasma-enhanced chemical vapor deposition (PECVD) provides a low-temperature, highly-efficient, and catalyst-free route to fabricate graphene materials by virtue of the unique properties of plasma. In this paper, we conduct reactive molecular dynamics simulations to theoretically study the detailed growth process of graphene by PECVD at the atomic scale. Hydrocarbon radicals with different carbon/hydrogen (C/H) ratios are employed as dissociated precursors in the plasma environment during the growth process. The simulation results show that hydrogen content in the precursors significantly affects the growth behavior and critical properties of graphene. The highest number of hexagonal carbon rings formed in the graphene sheets, which is an indicator of their quality, is achieved for a C/H ratio of 1:1 in the precursors. Moreover, increasing the content of hydrogen in the precursors is shown to reduce the growth rate of carbon clusters, and prevent the formation of curved carbon structures during the growth process. The findings provide a detailed understanding of the fundamental mechanisms regarding the effects of hydrogen on the growth of graphene in a PECVD process.

# Keywords

graphene growth | molecular dynamics | plasma-enhanced chemical vapor deposition | hexagonal rings

# INTRODUCTION

Since its first discovery in 2004, graphene, a two-dimensional (2D) single-atom-thick carbon sheet, has attracted tremendous attention because of its outstanding properties and potential use in many applications [1]. The unique hexagonal $sp^2$-hybridized carbon network of graphene results in outstanding electronic, mechanical, optical, and thermal properties [1-5]. To date, several approaches have been adopted to fabricate graphene, including mechanical exfoliation [1], chemical exfoliation [6], chemical synthesis [7, 8], epitaxial growth [9], and chemical vapor deposition (CVD) [10-12]. In particular, plasma-enhanced chemical vapor deposition (PECVD) has emerged as a promising new route to fabricate graphene with unique properties in a controllable, low-temperature, highly-efficient, and catalyst-free manner for various applications such as supercapacitors [11, 13], photocatalysis [14], transistors [15], sensors [16, 17], and solar vapor generation [18, 19].

During PECVD growth, factors such as plasma power, pressure, precursor composition, growth time, and substrate type can significantly affect the properties of graphene (e.g., morphology and quality) and its performance in practical applications [10, 20]. Among these factors, hydrogen has been experimentally demonstrated to play a crucial role in determining the properties of the PECVD-grown graphene [10, 21-25]. For instance, hydrogen in the precursors can accelerate the dissociation of carbon sources to carbon radicals by plasma [21, 25]. These carbon radicals will further react to form the final products [26]. Moreover, during the growth, hydrogen plasma can decrease the growth rates of graphene structures by etching carbon atoms at the boundaries, thus preventing the formation of amorphous carbon caused by rapid growth of carbon clusters [21, 25]. Hydrogen plasma has also been proposed for use in post-treatment processes to eliminate defects and promote the quality of graphene because of the strong etching effect of the plasma on $sp^2$-hybridized carbon atoms [24, 27]. However, the fundamental mechanisms governing the effect of hydrogen on the detailed growth behavior and properties such as quality of graphene remain elusive. Understanding this effect at the atomic scale is essential to controlling the growth of graphene by PECVD for use in various applications.

Reactive molecular dynamics (RMD) has become a widely used tool to reveal chemical reaction mechanisms at the atomic scale for applications such as catalysis [28, 29], etching [30], oxidation [31], pyrolysis [32], and nanomaterials synthesis [33-35]. Different from conventional MD, which describes interactions between atoms through pre-fixed bonds and charges, RMD simulates chemical reactions based on reactive force fields, where the breaking and formation of bonds are determined by the bond orders between atoms. [36, 37]. During the PECVD process, carbon sources are dissociated by the plasma into several radical configurations, $CH_x$ ($x$=0, 1, 2, 3) [38, 39]. While directly simulating the plasma effect by MD simulations remains difficult, these known radicals generated in plasma can be used as the precursors to simulate the plasma-enhanced growth process by RMD [40, 41]. To date, few papers are reported on RMD simulations of the graphene growth by PECVD, including the structural evolution of graphene during the growth process [41, 42]. Nevertheless, the effect of hydrogen on the detailed growth process and critical properties (e.g., quality) of graphene grown by PECVD remains unknown and warrants in-depth investigation.

In this work, we conduct RMD simulations to simulate the growth of graphene by PECVD. Carbon radicals are employed as dissociated precursors to mimic realistic PECVD growth processes. The C/H ratio of the precursors is investigated to elucidate the effect of hydrogen on the detailed growth process as well as quality and morphology of graphene. The corresponding detailed evolutions of the graphene formation processes are visualized to understand the fundamental mechanisms of the nanomaterial growth at the atomic scale. The results obtained from this work provide important insights into understanding graphene growth mechanisms by PECVD, and can also serve as guidelines for choosing precursors with suitable C/H ratios to fabricate graphene with desired properties for various applications.

## METHODS AND SIMULATION MODEL

Simulations are performed using the large-scale atomic/molecular massively parallel simulator (LAMMPS) [43]. The ReaxFF potential, which has been extensively proved to describe accurate reactions in hydrocarbon systems [44-46], is used to model the graphene growth process in this work. $CH_3$, $CH_2$, CH, and C radicals are employed as the dissociated precursors in the plasma to mimic actual PECVD growth processes [38, 39, 47, 48]. Simulations are conducted at a constant temperature of 3500 K in accordance with prior work wherein ReaxFF has been successfully demonstrated in hydrocarbon systems at temperatures ranging from 1000 to 3500 K [41, 49, 50]. For each C/H ratio, the amount of carbon in the precursors is kept constant, and the sizes of the simulation domains are modified to maintain the density of system at 0.2 $g/cm^3$. The detailed compositions of these radicals are shown in Table 1. Note that this constant density is much higher than the standard densities of typical carbon sources (e.g., methane, density 0.000717 $g/cm^3$ [41]). Such a high value of temperature and density is widely adopted in previous RMD simulations related to materials growth to increase the reaction rate and reduce computational costs to a realistic level [41, 51].

Canonical ensemble (i.e., NVT) is employed to simulate the growth process of graphene clusters, and temperature in the system is controlled by the Nose-Hoover thermostat method. Periodic boundary conditions are applied to the system to mimic the practical growth process. A time step of 0.1 fs is used to capture the details of reactions at high temperatures. The velocities and positions of carbon and hydrogen atoms are recorded every 10,000 steps (i.e., every 1 ps). The radicals are initially equilibrated under NVT at 300 K for 1 ns to achieve random dispersion within the simulation domain. Equilibration is followed by growth under NVT at 3500 K for 10 ns. The growth process is visualized using Open Visualization Tool (OVITO).

Table 1 Components of the precursors in the systems with different C/H ratios.

|  | Number of C | Number of CH | Number of $CH_2$ | Number of $CH_3$ |
|---|---|---|---|---|
| **Pure C** | 450 | 0 | 0 | 0 |
| **C:H=2:1** | 225 | 150 | 75 | 0 |
| **C:H=1:1** | 150 | 150 | 150 | 0 |
| **C:H=1:2** | 0 | 150 | 150 | 150 |

## RESULTS AND DISCUSSION

The simulated growth process of graphene at 3500 K based on precursors with a C/H ratio of 1:1 is shown in Fig. 1, in which the red spheres represent carbon atoms, and the blue spheres represent hydrogen atoms. Evolution of carbon structures for growth times ranging from 0 to 10 ns can be clearly observed. Initially, $CH_x$ radicals are dispersed uniformly in the simulation box. As the simulation proceeds, several long carbon chains appear and are subsequently cyclized to form polycyclic hydrocarbon clusters. These isolated polycyclic clusters are further diffused and merged to form a complete graphene sheet. Animation of this dynamic PECVD process of the graphene growth can be visualized in the Supplementary video.

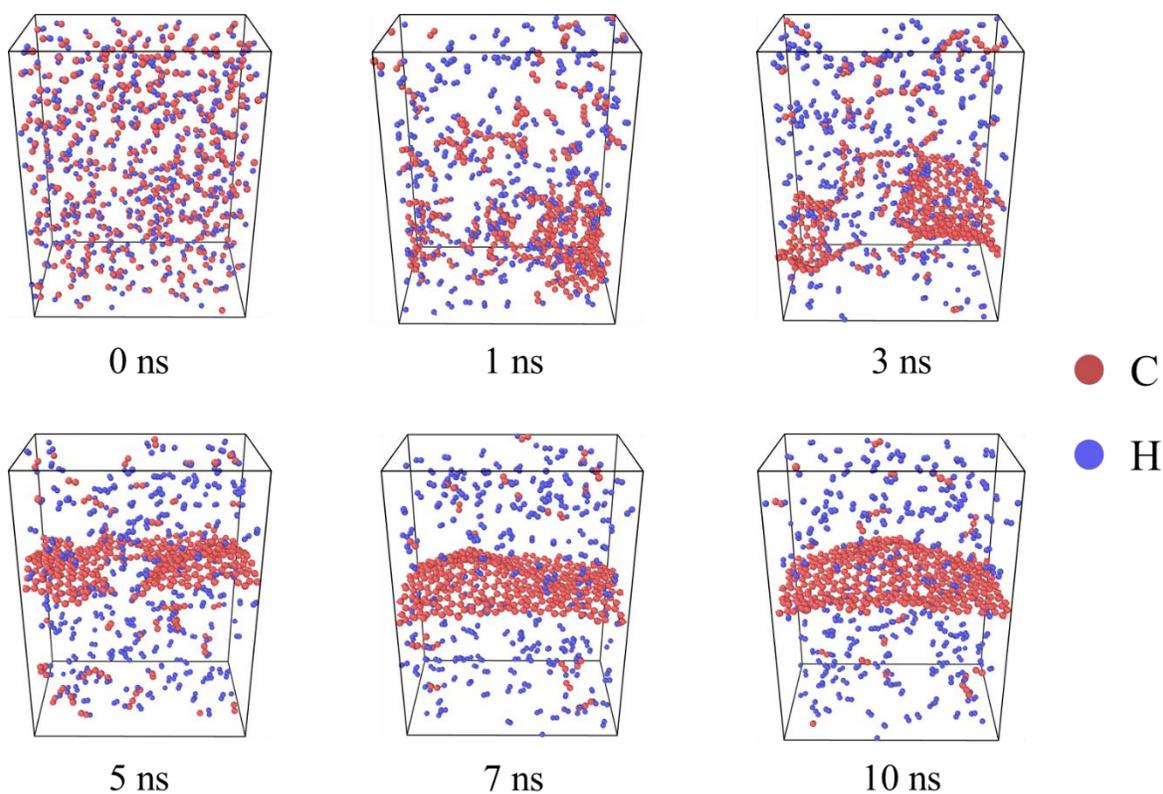

Fig. 1. Snapshots of representative dynamic processes of graphene growth simulated in MD for a C/H ratio of 1:1. Red spheres represent carbon atoms, and blue spheres represent hydrogen atoms.

A detailed, visual evolutionary depiction of the graphene network formation process with C/H ratio of 1:1 is shown in Fig. 2, wherein only the carbon atoms (red spheres) are shown for better visualization of the formed structures. At 0 ns, carbon atoms are dispersed uniformly, illustrating that radicals are in random positions after the initial equilibrium process process (Fig. 2a). Several short carbon chains are formed within the first 0.05 ns (Fig. 2b) as a result of reactions between hydrocarbon radicals [41]. These short carbon chains are subsequently and rapidly elongated to form longer carbon chains (Fig. 2c). Notably, a large number of branches are visible in the long chains. Cyclization of branches in the carbon chain has been reported to produce small carbon rings [35, 41]. In this simulation, branches on the long chains are indeed cyclized to produce polycyclic carbon rings, leading to the formation of graphene nuclei (Fig. 2d). Adjacent graphene nuclei then merge to form networks through cyclization of side chains (Figs. 2e-f). The graphene network continues to evolve by condensation of isolated short carbon chains at the periphery, and finally a complete graphene sheet is formed (Figs. 2g-i).

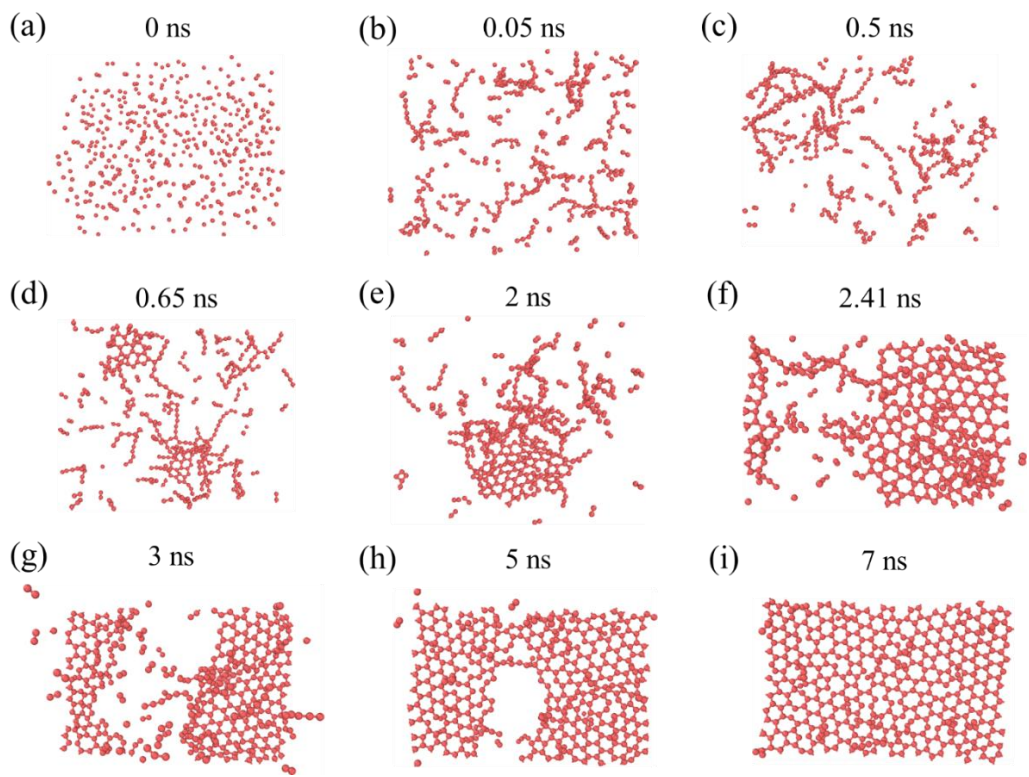

Fig. 2. Snapshots of the formation of graphene network for a C/H ratio of 1:1. Red spheres represent carbon atoms. Hydrogen atoms are not shown here for better visualization of the formed carbon structures.

Simulations based on hydrocarbon radicals with different hydrogen contents in the precursors (corresponding to different C/H ratios) are conducted to investigate the effect of hydrogen on the quality of the obtained graphene sheets, which is quantified by the average number of polygonal carbon rings, during the dynamic growth process from 0 to 10 ns (Fig. 3). As shown in Fig. 3a, in the system without hydrogen atoms (i.e., pure carbon system), carbon clusters agglomerate and curve severely to form a tube-like structure resembling those of carbon nanotubes (more details are shown in Supplementary Fig. S1). As the precursor content of hydrogen is increased, agglomeration and curving of carbon structures are observed to weaken. Moreover, in the systems with lower hydrogen contents (C/H=2:1 and 1:1), the carbon atoms tend to form a 2D network, indicating that the presence of hydrogen atoms contributes to the formation of a relatively flat graphene sheet rather than a curved structure caused by agglomeration and closing. The number of hexagonal carbon rings formed during the growth is examined to quantitatively evaluate the quality of the carbon structures for the different precursor C/H ratios (Fig. 3b). These results show that the growth rate of hexagonal carbon rings decreases when the hydrogen content in the precursor increases. In addition, a greater number of hexagonal rings is formed in the system with C/H ratio of 1:1 compared to that with C/H ratio of 2:1, which reveals a higher resulting graphene quality for the ratio of 1:1. Note that the formation of long carbon chains within 10 ns only occurs for a C/H ratio of 1:2, indicating that the growth rate of hexagonal carbon rings decreases significantly in systems with an excessive amount of hydrogen.

Pentagonal and heptagonal carbon rings are common defects in graphene structures [34]. The variations in the numbers of pentagonal and heptagonal carbon rings formed for different C/H ratios are shown in Fig. 3c and 3d, respectively. As the hydrogen content in the precursor increases, the formation of pentagonal and heptagonal carbon rings is retarded notably, which is similar to the trends observed for hexagonal carbon rings. More pentagonal and heptagonal rings are formed for a C/H ratio of 2:1 compared with other growth conditions, while the number of formed hexagonal carbon rings under this growth condition is smaller (Fig. 3b), indicating that the obtained graphene sheet contains more pentagonal and heptagonal ring structures as defects for a C/H ratio of 2:1. Interestingly, notable decrease of pentagonal and heptagonal rings during 5-7 ns can be observed for C/H ratio of 2:1, while the number of hexagonal rings increases. This phenomenon indicates that a healing process may occur where some pentagonal and heptagonal carbon rings are transformed to hexagonal rings because of carbon diffusion and structural reconstruction. Such a transformation process from pentagonal and heptagonal carbon rings to hexagonal rings was also observed in previously simulated thermal CVD growth processes [34, 52].

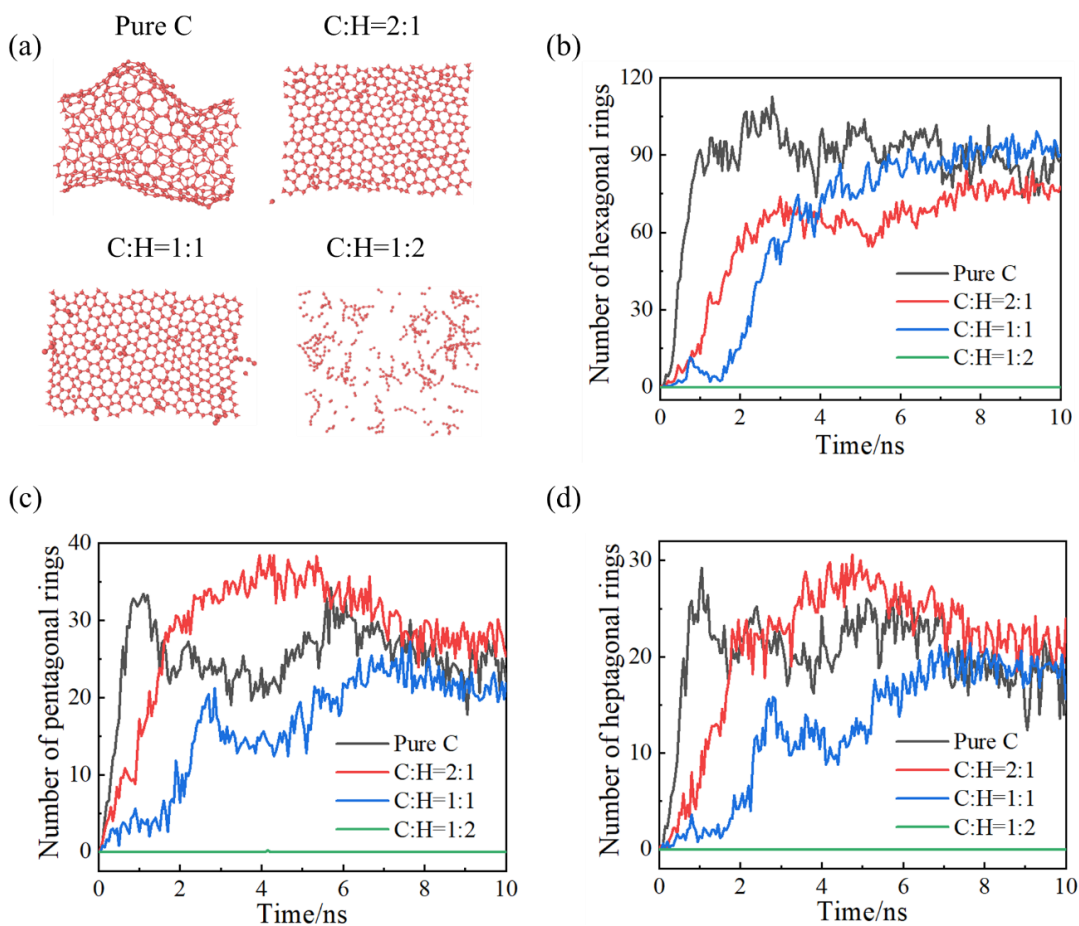

Fig. 3. (a) Snapshots of carbon structures formed at 10 ns for different C/H ratios. Number of (b) hexagonal carbon rings, (c) pentagonal carbon rings, and (d) heptagonal carbon rings as functions of time for different C/H ratios.

To illustrate the effect of hydrogen on the morphology of the obtained carbon structures, we further compare the simulated growth processes of the pure carbon system and the one with a C/H ratio of 1:1, as shown in Fig. 4. In the pure carbon system, since no hydrogen atoms exist in the simulation box, only C-C bonds can be formed at the periphery of the clusters (Figs. 4a-b). Chemical reactivity of carbon atoms at the edges is high because of their unsaturated states.

Therefore, the peripheries of small carbon clusters can be easily connected by cyclization of these carbon dangling bonds, leading to the agglomeration and closing of carbon structures, as shown in Figs. 4c. Finally, a tube-like carbon structure is formed (Fig. 4d and Supplementary Fig. S2). While in the system with a C/H ratio of 1:1, hydrogen atoms can form C-H bonds with the carbon atoms at the edges of carbon clusters, thus significantly decreasing the number of carbon dangling bonds (Figs. 4e-f). In this case, more carbon atoms at the periphery are saturated due to the existence of hydrogen atoms, thus weakening the cyclization of carbon dangling bonds and preventing graphene sheet from rolling into a closed structure (Fig. 4g). In this case, relatively flat graphene sheets are formed (Fig. 4h). However, the decreased number of carbon dangling bonds also leads to lower growth rates, which can explicitly explain the results shown in Fig. 3. These simulation results agree well with the experimental observation that hydrogen precursors in plasma transformed the carbon morphology from nanospheres to nanosheets [53].

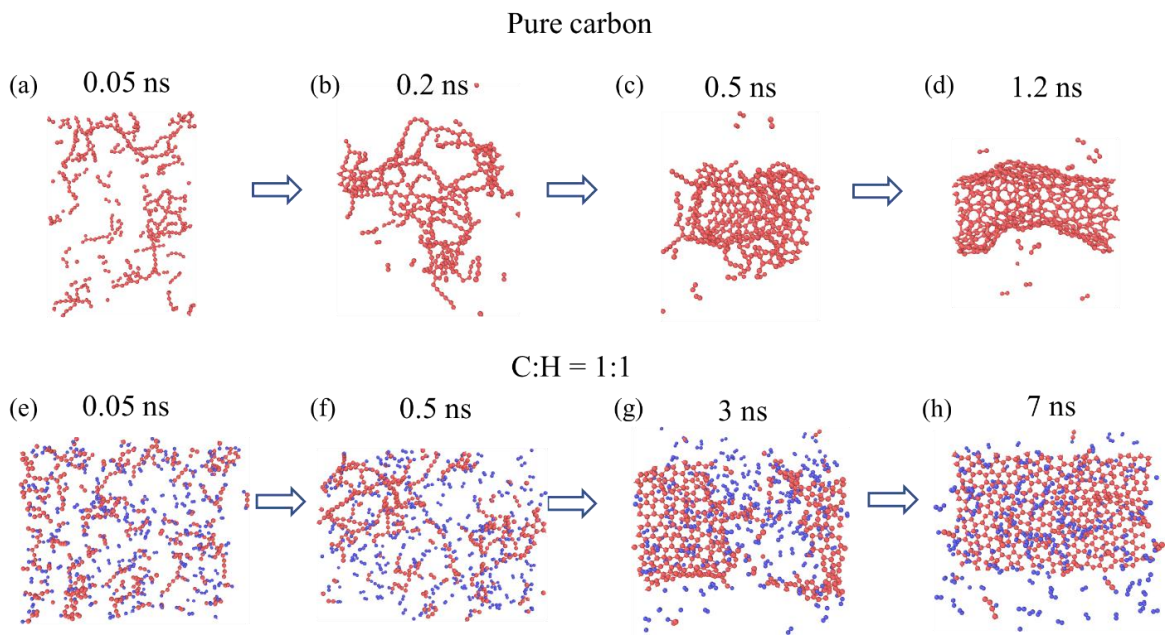

Fig. 4. Structural evolution of the pure carbon system (a) 0.05 ns, (b) 0.2 ns, (c) 0.5 ns, and (d) 1.2 ns. Structural evolution of the system with a C/H ratio of 1:1 (e) 0.05 ns, (f) 0.5 ns, (g) 3 ns, and (h) 7 ns. Red spheres represent carbon atoms, and blue spheres represent hydrogen atoms.

To further understand the effect of hydrogen on the formation behavior of carbon structures during growth, the number of carbon atoms in the largest carbon cluster is analyzed as a function of the growth time with different C/H ratios in the precursors, as shown in Fig. 5. In the pure carbon system, or the mixture systems with lower hydrogen contents, the number of carbon atoms increases gradually to a maximum value and then becomes relatively stable during the evolutionary process, corresponding to the continuous growth of carbon clusters in the initial stage and equilibrium state afterwards. Nonetheless, this phenomenon is not quite noticeable for the system with a high hydrogen content (C:H=1:2), which exhibits a relatively smaller size of the formed carbon networks. The growth rate of the largest carbon clusters is observed to decrease significantly as hydrogen content in the precursor increases. For instance, in the pure carbon system, the largest carbon cluster grows rapidly to reach a maximum size after approximately 1 ns, while for the largest carbon cluster in the system with a C/H ratio of 1:1, it takes approximately 7 ns to reach the maximum size (Fig. 5). These results further illustrate that hydrogen can efficiently prevent carbon clusters from rapidly growing to form curved structures, and an excessive amount of hydrogen in the precursor can greatly deter the growth of carbon structures.

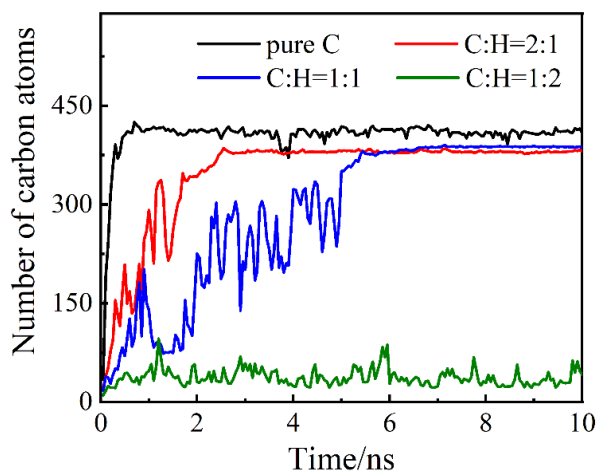

Fig. 5. Number of carbon atoms in the largest carbon cluster as a function of growth time for different C/H ratios in the precursors.

Additionally, hydrogen content in the precursor may also affect the distribution of carbon cluster sizes during the growth process. In this case, carbon clusters with different sizes (indicated by the number of carbon atoms) in systems with different C/H ratios are analyzed every 50 ps, as shown in Fig. 6. Fig. 6a evaluates the population of small-size carbon clusters (containing 4-10 carbon atoms) as a function of growth time. In the pure carbon system and the mixture systems with lower hydrogen contents (C/H=2:1 and 1:1), the number of small-size carbon clusters first increases due to the reactions between carbon radicals, and then decreases gradually to a minimum value because the small-size carbon clusters further grow to form larger ones. Notably, the reduction rate of the population of small-size carbon clusters decreases as the hydrogen content in the precursors increases, indicating that hydrogen deaccelerates the growth of small-size carbon clusters into larger ones. Similar effects of hydrogen content on the growth behavior of medium-size carbon clusters (containing 11-100 carbon atoms) are also observed in Fig. 6b. Therefore, the appearance of large-size carbon clusters (containing more than 100 carbon atoms) is postponed when the hydrogen content in the precursors increases, as shown in Fig. 6c. However, the numbers of both small-size and medium-size carbon clusters exhibit negligible changes during 7 ns in the mixture system with a higher hydrogen content (C/H=1:2), which can be attributed to the greatly weakened reactions caused by the existence of excessive hydrogen in the precursor. Interestingly, the hydrogen content in the precursor also affects the stability of the formed large-size carbon clusters (Fig. 6c). For instance, in the pure carbon system and the system with a low hydrogen content (C/H=2:1), large-size carbon clusters first appear at 0.05 ns and 0.25 ns, respectively, and remain stable thereafter. While in the system with a higher hydrogen content (C:H=1:1), the first large-size carbon cluster appears at 0.55 ns and is quickly dissociated to smaller carbon clusters, which are connected again to form a large-size carbon cluster at 0.7 ns. Such a process repeats several times until a stable, large-size carbon cluster is formed at 1.65 ns. In this case, we can infer that it may take longer time to obtain stable large-size carbon clusters in the system with a C/H ratio of 1:2.

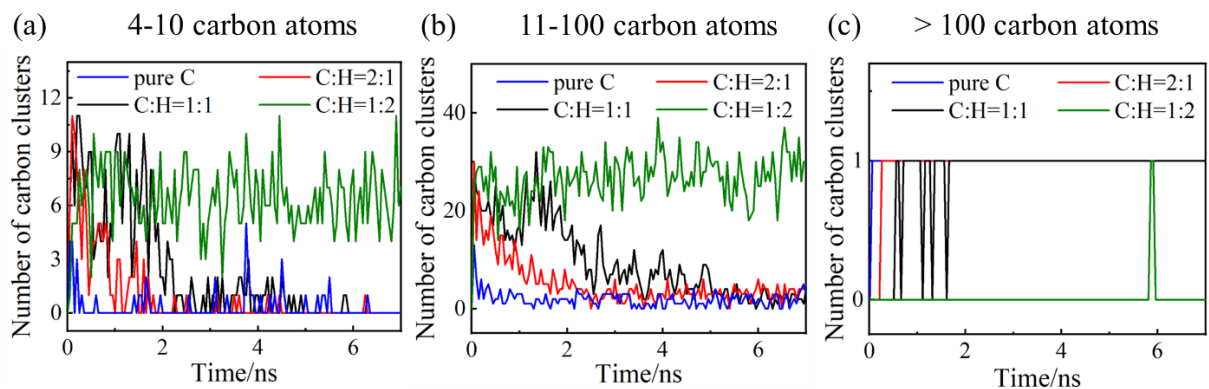

Fig. 6. Number of carbon clusters containing (a) 4-10 carbon atoms; (b) 11-100 carbon atoms; (c) >100 carbon atoms as a function of time for different C/H ratios in the precursors.

Dynamic change of potential energy of the entire system can reveal important information of the growth behavior of graphene in the process, which can be calculated using the ReaxFF reactive force field adopted in our simulation. Fig. 7 exhibits comparative potential energies of the pure carbon system and mixture systems with C/H ratios of 2:1 and 1:1 during the growth process. In each of the three systems, the potential energy is observed to decrease gradually until it reaches equilibrium. During the growth process, isolated carbon atoms are prone to reduce their potential energy through the formation of carbon chains, and these carbon chains will be further transformed to more stable carbon networks through cyclization, agglomeration and closing [34, 41]. Note that the reduction rate of potential energy in the initial growth stage before reaching equilibrium varies with the hydrogen content in the precursors. For instance, in the pure carbon system, the potential energy decreases rapidly and reaches equilibrium within only 1 ns, which can be attributed to the fast formation of carbon structures due to the high reactivity of dangling carbon bonds. Moreover, curling and closing of carbon networks will also contribute to the reduction of potential energy [41]. When the hydrogen content in the precursors increases, an increasing number of C-H bonds, instead of C-C bonds, are formed at the peripheries. The decreased number of carbon dangling bonds slows the growth of carbon networks and prevents the graphene sheets from rolling into curved structures, leading to a decreasing reduction rate of potential energy.

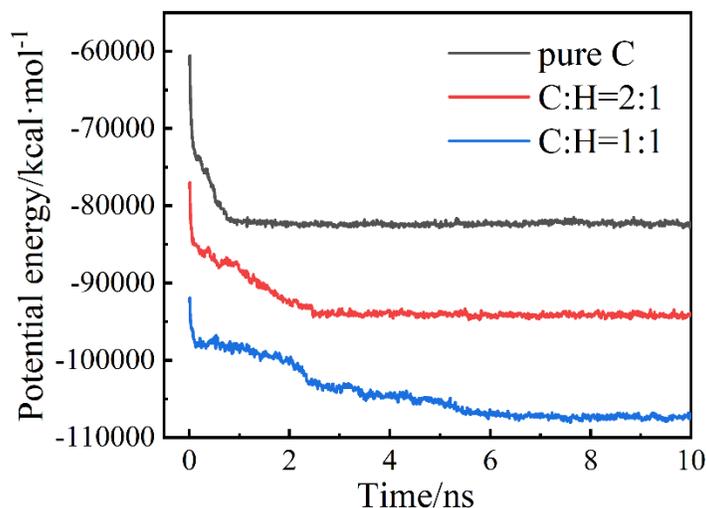

Fig. 7. Comparative potential energies of the systems with different C/H ratios in the precursors as a function of the growth time.

## CONCLUSION

In summary, RMD simulations using the ReaxFF potential have been conducted to provide an atomistic understanding of the plasma-assisted graphene growth using hydrocarbon precursors with different C/H ratios. The simulation results indicate that hydrogen content in precursors significantly affects the growth behavior and critical properties of graphene during PECVD. Increasing the hydrogen content in precursors facilitates the formation of C-H bonds at the periphery of carbon clusters, leading to a decreasing number of carbon dangling bonds. Accordingly, the growth of carbon clusters can be efficiently slowed by introducing hydrogen in the precursor, thus preventing graphene sheet from rapidly growing and rolling into a curved structure. Moreover, the existence of hydrogen is found to substantially affect the number of polygonal carbon rings and the quality of the graphene sheets. The quality of graphene is observed to be the highest for a C/H ratio of 1:1. Our work provides important insights into understanding the growth mechanisms of graphene by PECVD and helps guide the practical growth processes by optimizing the precursor components to fabricate well-controlled graphene towards various applications.

## Acknowledgements

G.X. and S.W. thank the University of Texas at Dallas startup fund, the ACS PRF (DNI60563) and NSF (Grant No. CBET-1937949, CBET-1949962 and CMMI-1923033). T.L. and D.H. thank the support from the NSF (Grant No. CBET-1937923 and CBET-1949910). The simulations are supported by Ganymede HPC Cluster at the University of Texas, TACC Lonestar5 system, the Notre Dame Center for Research Computing, NSF through XSEDE computing resources provided by TACC Stampede2 under grant number TG-CTS100078.

# Supplementary Materials for

# Molecular Understanding of the Effect of Hydrogen on Graphene Growth by Plasma-Enhanced Chemical Vapor Deposition

Shiwen Wu[1,+], Dezhao Huang[2,+], Haoliang Yu[1], Siyu Tian[1], Arif Malik[1], Tengfei Luo[2,*], Guoping Xiong[1,*]

[1.] Department of Mechanical Engineering, The University of Texas at Dallas, Richardson, Texas 75080, USA

[2.] Department of Aerospace and Mechanical Engineering, University of Notre Dame, Notre Dame, Indiana 46556, USA

[+] These authors contributed equally to this work.

* Corresponding authors: Tengfei Luo: tluo@nd.edu; Guoping Xiong: guoping.xiong@utdallas.edu

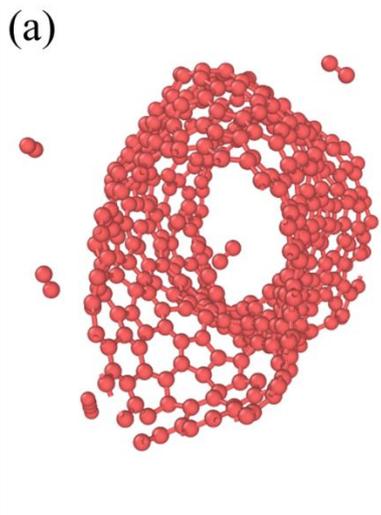 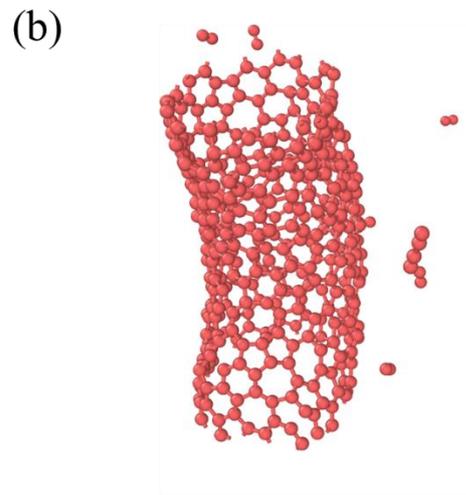

Fig. S1. Snapshots illustrating the tube-like carbon structure in the pure carbon system at 10 ns.

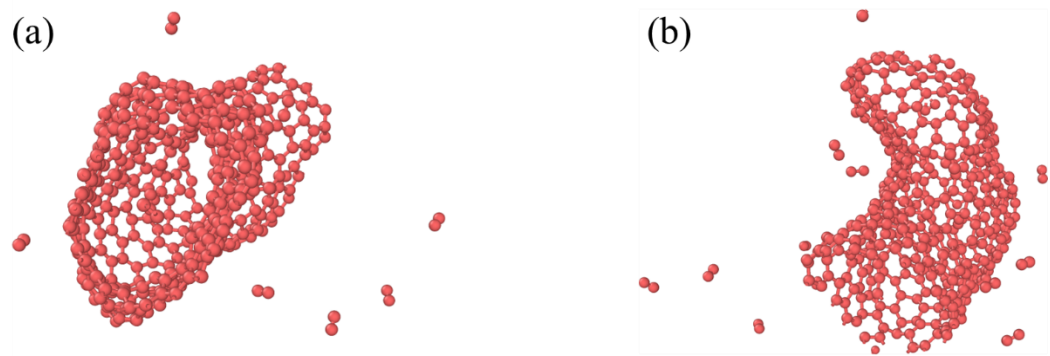

Fig. S2. Snapshots illustrating the tube-like carbon structure in the pure carbon system at 1.2 ns.